# Comment on 'Angles in the SI: treating the radian as an independent, unhidden unit does not require the redefinition of the term "frequency" or the unit hertz'

Petr Křen

Czech Metrology Institute, Okružní 31, 63800 Brno, Czech Republic
E-mail: pkren@cmi.cz

**Abstract**

In the paper by P. Quincey (2020) [1], the author claims that angles are not dimensionless quantities and that the radian should be a new independent base unit. However, the claim is not supported and the proposed redefinition will cause several problems not only in physics, but also in mathematics. The proposed "unhiding" of the radian will bring hidden problems not discussed in detail. This includes e.g. an introduction of quantities of different kind for time and the corresponding units defined as the second per radian.

## 1. Introduction

The quantity equation is generally based on the property that *all* units are hidden, including the units for the plane angle. For example, Ohm's law can be written as the quantity equation $U = RI$ because all quantities $q$ consist of the numerical value $\{q\}$ and the unit $[q]$. However, the numerical value equation needs a unit specification to avoid a discrepancy in the results based on particular units. For example: $\{U\}_V = \{R\}_\Omega \{I\}_A = 1000\{R\}_{k\Omega}\{I\}_A$.

The quantity named the plane angle is traditionally treated as a dimensionless ratio. The unit of plane angle $\theta$ expressed in terms of base units is m/m according to the 9th edition of the SI Brochure [2]. The dimension of the plane angle in dimensional analysis for the seven base quantities is given as dim $\theta$ = $T^0 L^0 M^0 I^0 \Theta^0 N^0 J^0 = 1$ with all powers equal to zero. The plane angle is traditionally given by the *quantity* equation

$$\theta = \frac{s}{r}, \tag{1}$$

where $s$ is the length of the circular arc and $r$ is the length of the radius of the circle. The dimensionless ratio of these two lengths can be used directly as an argument of the trigonometric functions such as

$$\sin(\theta) = \sin\left(\frac{s}{r}\right). \tag{2}$$

Note that all non-linear mathematical functions must have dimensionless arguments. Also note that we can use only the right-hand side of equation (2) and we do not need to define the plane angle at all. The plane angle is redundant information in the Cartesian coordinates. Everything, what we can describe in the polar coordinates (with an angle), we can describe in the Cartesian coordinates (without an angle). Thus, the plane angle is neither fundamental (with the radian as a base unit) nor independent.

By analogy, the efficiency $\eta$ is, for example, a ratio of energies and it can be used in the logarithm function as

$$\log(\eta) = \log\left(\frac{E}{E_0}\right) \tag{3}$$

or these ratios can be used directly without a definition of the efficiency $\eta$.

It is suggested in [3] (cited in [1] as reference 7) that angles are not inherently length ratios that, however, does not imply that angles are not dimensionless. The angle can be alternatively defined using the area of circular sector $A$ by the quantity equation as

$$\theta = \frac{A}{r^2/2}. \tag{4}$$

However, this does not change anything on its dimensionality given by equation dim $\theta$ = L$^{2-2}$ = L$^0$ = 1 even if we use for the definition a mass ratio of the sector from e.g. a sliced pie with dim $\theta$ = M$^{1-1}$ = M$^0$ = 1. In [3] the authors also suggest to use a natural constant *unit* named $\theta_N$ = [$\theta$] = rad and the numerical value $\{\theta\}$ = $s/r$ in equations as

$$\{\theta\} = \frac{\theta}{[\theta]} = \frac{\theta}{\theta_N} = \frac{s}{r} = \frac{A}{r^2/2}. \tag{5}$$

Note that the equation is not a quantity equation and the authors of [3] do not provide instructions on how to write pure quantity equations with angles. Also note that the area of a circular sector $A$ cannot contain angular units because the area of a circular segment is lower by the area of the triangle that is generally without units like the radian and we cannot subtract quantities of different kinds. However, the suggestion in [3] introduces a necessity to write e.g. sin ($\{\theta\}_{rad}$) = sin ($\theta/\theta_N$) = sin ($\theta$/rad) always for any angle instead of the common simple quantity equation sin ($\theta$).

By analogy, the suggestion is equivalent to a necessity to write log ($\{\eta\}_1$) = log ($\eta/1$) to avoid efficiencies expressed in percentages like log ($\{\eta\}_\%/100$) = log ($\eta/100\ \%$). The corresponding numerical equation is

$$\{\eta\} = \frac{\eta}{[\eta]} = \frac{\eta}{\eta_N} = \frac{E}{E_0} \tag{6}$$

with the natural efficiency $\eta_N$ equal to 1 or to 100 %. Of course, the efficiency is not inherently energy ratio (e.g. it can be a power ratio) and that also does not imply that efficiencies are not dimensionless. Taking into account that conventionally and traditionally rad = 1, we can see that dim $\theta$ = dim $\eta$ = 1. We do not need the angle $\theta_N$ = 1 nor the angle of one revolution $\theta_{rev}$ = 2π$\theta_N$ = 2π rad = 2π introduced in [4] (cited in [1] as reference 6). Moreover, it is not clear throughout these papers whether the symbol $\theta_N$ represents the angle (a quantity written in italics) or the unit (that is equal to the radian). However, it does not have a practical consequence for numerical results because it is equal to one.

## 2. The frequency and time

The ISO standard 80000-3:2019 defines the duration $t$ as "measure of the time difference between two events", whereas the period $T$ as "duration of one cycle of a periodic event". That is, the period is a specific kind of the duration. The frequency $v$ is defined as "inverse of period duration", and thus it is restricted only to periodic events. However, the frequency can be generally a function of time $v(t)$ because it is not perfectly stable. Thus, it should be allowed to be used for non-harmonic phenomena, the frequency modulation, a counting of events with the Poisson distribution, the frequency noise, the Earth's rotation frequency, etc. The ISO definition restricts the frequency only to events with a spectrum of the Dirac delta function, i.e. spectra with zero linewidth that are not present in the real metrological world. It must be noted that the definition of the second is (and will be) based on atomic events that are also *not regular* in time and the spectrum of an atomic transition does not have the normal distribution. Such spectral distributions (as well as the Cauchy distribution which is called the Lorentz distribution in physics) do not fulfil the central limit theorem and the mean value is rigorously undefined for them. The previous definition of the second was chosen as "the *duration* of 9192631770 *periods* of radiation corresponding to the *transition*", where the link between the harmonic periods and non-harmonic transitions was done by the word "corresponding". The new SI definition [2] is "the caesium frequency ... *transition frequency* ... to be 9192631770 when expressed in the unit Hz, which is *equal* to s$^{-1}$" without an explicit mention of "an average" or a harmonic frequency. The equality (that is a symmetric relation by its own definition) between Hz and s$^{-1}$ suggests that the definition introduces the frequency also for non-periodic phenomena. However, it is recommended (also in [2]) that the frequency "always be given explicit units of Hz" and the reciprocal second s$^{-1}$ should not be used (even though it is equivalent by definition).

The ISO standard 80000-3:2019 defines the angular frequency $\omega$ as "rate of change of the phase angle" given by the quantity equation $\omega = 2\pi\nu$, and thus the definition restricts it to harmonic angular frequencies. The ISO standard does not explicitly mention what the factor "$2\pi$" means. The number $\pi$ is generally a mathematical constant and it is not a quantity that should be written using italic type. Thus, the dimension of $\omega$ is equal the dimension of $\nu$. In the paper [1], its author promotes (as well as in its previous papers) "the constant ... $\theta_N$" that is equal to "1 rad" to generate "the complete equation" $\omega = 2\pi\theta_N\nu$. That means, a something must be placed into the quantity equation (as well as the radian to equations for units $2\pi\cdot$rad$\cdot$Hz) instead of the current simpler notation $\omega = 2\pi\nu$ and $2\pi\cdot$Hz, respectively. However, the $2\pi\theta_N$ or $2\pi\cdot$rad cannot be an angle of the full cycle together with the statement that the plane angle has its own dimension because the Taylor series of trigonometric functions with the plane angle in their argument will be additions of terms with different dimensions (powers of radians) that is generally not allowed in physics.

The quantity $\omega$ allows to simplify the writing of arguments in the trigonometric functions (with the unit of period equal to the constant $2\pi$), where their argument $2\pi\nu t = \omega t$ gives the same results for each temporal period $T$. For example, the period $T = 0.02$ s and the frequency $\nu = 50$ Hz lead to the argument equal to $2\pi$ (the period of trigonometric functions). The angular frequency can be calculated as $\omega \approx 314$ Hz by using the current "Radian Convention" (the recent term that is probably coined by the author of [1]). The angular frequency $\omega$ corresponds to the number of radians turned per second (a part of cycle per second). That is, the radian reached by a rotation within elapsed time is the angular period given by equation $P = 1/\omega \approx 0.0032$ s for the mentioned example, whereas in the equation corresponding to proposals in [1] time will be equal to $P \approx 0.0032$ s/rad. Note that, for example, the hour was exactly defined as the time necessary for a rotation of the Earth by a given angular fraction of the full cycle and 1-hour time zones are 15 degrees wide (the angular period 240 s/°). The Earth's rotation is very stable due to the conservation of the angular momentum $L$ that is connected with the plane angle [4]. Time for a rotation has traditionally a unit for the same kind of quantity as the angular period $P$. However, time or frequency generally do not have different units from the angular time and the angular frequency, respectively. But the redefinitions proposed in [1] introducing a dimension for the plane angle will also change these deeply rooted units. Universal Time (UT, where the angular-time second is based on an angular fraction of the whole cycle rotated by the Earth in time) will be a *different quantity of different dimensions* from Coordinated Universal Time (UTC, where the second is time obtained from the frequency of events) as a consequence of the proposal in [1]. However, we do not need specific units like "second per radian" or "second per particle event" as well as "radian per second" or "particle event per second". These quantity specifications generally should not be included in units, otherwise we will also need to define new units with their own "independent" dimensions such as "production of apples or oranges per second", "seconds needed for a production of something", "person-hour" etc., even though tendency for them is large. In the past, every profession had its own units and it is the great success of the SI that the number of units was reduced and new units must be introduced or upgraded very carefully.

The same problem occurs in the spatial domain for the wavelength $\lambda = c/\nu$ by the definition of the metre using the speed of light $c$. The argument for trigonometric functions with spatial frequencies is $2\pi/\lambda\cdot x = kx$. That is, the wavelength in the spatial domain (the period $T$ in time domain) corresponds directly (without the $2\pi$ factor) to the wavenumber (the frequency $\nu$, respectively). And also the angular wavelength (the angular period $P$) corresponds directly to the angular wavenumber (the angular frequency $\omega$, respectively). However, the second half of these quantities will be affected by the proposed redefinition. Note that the metre was defined as number 1 650 763.73 of *wavelengths* (not in wavenumbers) of the transition in krypton (also by using the angular fraction of wavelengths in interferometers). The numerical value of the speed of light was then measured as $c = \lambda\nu$. Moreover, the first definition of the metre (defined as the angular fraction of "circumference" and realized by the grade/arc measurements of the Earth, i.e. in units like m/rad or m/° using an analogy according to [1]) was replaced by the current definition with the length in meters. That means, there was also a historical transition between the quantities according to the interpretation based on [1] which suggested the plane angle to have its own dimension.

It must be noted that the vector product, such as $\boldsymbol{L} = \boldsymbol{r} \times \boldsymbol{p}$ mentioned in [1], denoted by the symbol ×, is generally not a linear function of the plane angle or its power. The dimensionless factor of the vector products is generally proportional to the sine function result of the angle between the vectors in its argument, and thus the result (such as the angular momentum) cannot have a unit like the radian because the plane angle is *not independent* of the space (e.g. length). However, the proposal in [1] suggests to modify the definition of the angular momentum by dividing the quantity equation using the plane angle $\theta_N$ (Is this "complete equation" still a quantity equation?). This way, the "unhiding" unit of the radian also needs a modification of the quantity equation. The consequent redefinitions also include e.g. the moment of force $\boldsymbol{\tau} = \boldsymbol{r} \times \boldsymbol{F}$ as well as the angular velocity $\boldsymbol{\omega} = \boldsymbol{r} \times \boldsymbol{v} / r^2$ (with the same units as the angular frequency $\omega$ according to e.g. the ISO standard) that are both related to the angular momentum.

## 3. Discussion

The *ambiguity* of the radian as a unit was already mentioned in 1917 [5]. The dimension one of the plane angle corresponds to the fact that the length of some curved line is the same as the length of straight line. These mathematical objects have the *same dimension*, and thus there is no place for a dimension of the plane angle defined as their ratio. Further, in [1] the author states that "it does mean that dimensional analysis cannot be applied". However, the VIM [6] declares "quantity of dimension one ... quantity for which all the exponents of the factors corresponding to the base quantities in its quantity dimension are zero ... the ratios of two quantities of the same kind. EXAMPLES Plane angle, solid angle, refractive index...".

According to [1], the definition of a "constant" is $\theta_N = 1$ rad, or simply $\theta_N =$ rad. And it is, according to "the Radian Convention", equal to 1, and thus it is conventionally omitted in mathematics (that also deals with the plane angle) as well as any other "1". In the paper [7] that is cited in [1] (as reference 1), its authors must also introduce a completely new physical quantity that is named "the angular radius of curvature" to introduce the radian as a unit. In [7] this has a consequence that the unit for the moment of inertia contains rad$^{-2}$. However, the second moments are mathematically general (e.g. used for calculations of the standard deviations of arbitrary variables) and they are also used for calculations of the multipole moments of electric charges and not only for the moment of inertia, and thus they cannot contain the radian as well [8]. In my opinion, a full list of quantities is needed for a comparison to know, how many quantities will be affected by such a proposal. However, such a list has never been shown, and thus it is hidden to readers. In [1] The the author wants to "unhide" the unit "rad" as well as the constant [quantity?] "$\theta_N$", and thus to rewrite various quantity equations. This seems to be very complicated and problems are hidden to readers, whereas it is only stated in the conclusions that "the standard definitions of 'frequency' and the hertz should remain unchanged" or "the actual changes required would be small" without showing more examples of changes that will be needed (some of them are shown in [7]) as the consequence.

The SI Brochure [2] only mentions that the plane angle has the radian as a special unit name and that rad = m/m is expressed in terms of base units. That means, the radian is a special name for one in the case of the plane angle. The ISO standard 80000-3:2019 calls the plane angle only "angular measure", defined as the ratio $s/r$, and allows both alternatives for its units: "rad" and "1". All ratios of quantities of the same kind are quantities of dimension one. Thus, units of such ratios lose information about units of quantities that were cancelled out. It is the general property of reducing fractions and it cannot "lead to algebraic nonsense" claimed in [1]. Therefore, it is not necessary to define any special unit for a ratio of quantities of the same kind, including the units for the plane angle and the solid angle. However, they "have the option of being expressed with units (m/m, mol/mol)" according to [2]. Thus, the "rad" can also be replaced by "m/m" and the radian may not be defined at all. Moreover, the suggested adoption of the radian as a base unit means that (up to now) something dimensionless will have its own dimension.

The system with five base units was proposed in [4]. It is based on four conserved quantities (the mass-energy, the linear momentum, the angular momentum and the electric charge) and the action. The reason is that conserved *quantities* are "independent". It must be noted that e.g. the colour charge, the weak isospin, the lepton number and the total probability (equal to 1) are also conserved quantities that were not taken into account in the proposal in [4] which seems to be incomplete. However, the question is whether their *units* also must be independent. The new SI system linked the base units to the second using the fixed numerical factors in their definitions. Time is the quantity which is measured the most precisely, further, a duration of time is essentially needed to perform any measurement and the measurements can be reduced to it using these SI unit factors. The traceability of units *depends* on the factors rather than being *independent*. In [1], the title contains a claim that "the radian as an independent, unhidden unit", but the author does not bring any relevant argument for independence of the angle and why we need to unhide "$\theta_N$" and "rad". The units are generally dependent by conventions and they are generally hidden in all quantity equations. Moreover, the plane angle is dependent by its definition, namely it is a ratio of quantities of the same kind.

I strongly disagree with an introduction of the radian as the base unit of the International System of Units (SI). In 2016, the president of the Consultative Committee for Units (CCU) asked the Consultative Committee for Length (CCL) about its opinion. The report of CCL was created and based on several contributions (including mine). The *consensus* of CCL stated that there was no support for the proposal to make the radian a *base* unit of the SI. It is also a historic recurrence, when we take into account that rad = 1. In 1998, the CCU recommended an adoption of a new unit named "uno" representing 1 in dimensionless quantities that will be denoted as U = 1. It can overcome the conventional restriction (this is only a conventional problem and not a physical one) to use the prefixes only with some symbol of a unit (also like e.g. the microradian that can be, however, replaced by µm/m). However, it has not been adopted by the International Committee for Weights and Measures (CIPM) until now. Moreover, it is suggested in [1] that a unit equal to one (the radian) should have its own dimension (being the base unit).

The radian and the steradian were formerly (only two) SI *supplementary* units, but this category has been abolished in 1995, approving the CIPM recommendation U1 (1980) that they are *d'unités sans dimension*. Already at the CCU meeting in 1980, the document 80-6 presented a suggestion introducing a dimension to the plane angle by equation $\alpha = \alpha_0 \cdot s/r$, where $\alpha_0$ is effectively the same as $\theta_N$ and it can produce a base unit. Thus, the proposal of the author of [1] is not a new; moreover, it was already (except by two delegates) unaccepted in the previous century with a good reason and with the conclusion that supplementary units are dimensionless. In [1] the author adopts that the steradian should be a *derived* unit equal to $rad^2$ instead of the long-established practice in mathematics, where $sr = 1 = 1^2 = rad^2 = rad$ automatically. Further, the author mentions that if the base unit will be "considered too radical", he proposes a "new category" named the *complementary* unit that should be created for the radian. However, the radian is still equal to 1, and therefore this is an unnecessary categorisation without any metrological added value.

Time was commonly measured by pendulum clocks. The motion of a simple ideal pendulum swinging at the plane angle $\theta$ is given by equation

$$\frac{d^2\theta}{dt^2} + \omega_0^2 \sin\theta = 0 \tag{7}$$

where the angular frequency of swings $\omega_0$ for the small-angle approximation is given by

$$\omega_0 = \sqrt{\frac{g}{l}} \tag{8}$$

where $g$ is the acceleration of gravity and $l$ is the length of the pendulum. The approximative solution for small amplitudes $\theta_0$ can be written as

$$\theta(t) \approx \theta_0 \sin(\omega_0 t). \tag{9}$$

We can see that the angular frequency in not necessarily connected with a real full revolution. The numerical factor $2\pi$ in the definition $\omega = 2\pi\nu$ does not corresponds to the real plane angle. The phase angle (defined in the ISO standard) is "angular measure" in the complex plane with an imaginary axis. Further, we can see that it is not easy to make the dimensional analysis for equation (7) if the radian has its own dimension. Again, a very complicated notations of equations will be needed. The solution is that all $\theta$ and $\omega_0$ will be divided by $\theta_N$. This is a general property of the proposal in [1]. The "Radian Convention" definition is $\theta = s/r$, whereas the proposal divides it further as $\Theta = \theta/\theta_N$ (as well as $\omega_0/\theta_N = 2\pi\nu_0$) to be surely dimensionless, that is, however, only an extra work. The redefinition of the radian in a way that "a right angle equals $\pi/2$ rad" as proposed in [1], effectively equivalent to definitions like "a right angle equals $\pi/2$" or "a complete cycle equals $2\pi$", cannot bring an independence of other SI base units (How to define what is the right angle without them to avoid a tautology?). The measurement of the plane angle will need some of them. The *mises en pratique* are prepared for all base quantities. I cannot imagine, how it can be prepared also for the radian independently of other SI base units.

In [1], the author's fear of confusion between $\omega$ and $\nu$ (based on the same units that allegedly imply the statement "$\omega = \nu$") is not a relevant argument. Moreover, it is not necessary to introduce the angular frequency at all (as well as the plane angle), everything related can be expressed by the frequency. The specifying information is generally "hidden" in the definition of *quantities* (and it should not be transferred into the units) and it is also valid for dimensionless ratios such as the efficiency (the details about the efficiency must be specified in the quantity definition and a new unit "one" should not be defined). The common practice is "hiding" numerical factors within all quantities such the radius $r$, the diameter $d = 2r$, the circumference $c = 2\pi r$ etc. (without a fear of confusion between them due to the same length units) rather than introducing units like the radial meter, the diametral meter, the circumferal meter etc. in order to "avoid" strange statements like $r = d = c$. The mathematical constant $2\pi$ (established well before the existence of the Metre Convention, its symbol being derived from circumference, περιφέρεια in the Greek, and not from an angle) does not need its physical unit because it has the mathematical origin. The same argument of the mathematical constant $2\pi$ applies also to the Planck constant $h$ and the definition of its reduced version $\hbar$ that is also not necessary to be introduced because, for example, we can use the substitution $h\nu = \hbar\omega$.

## 4. Conclusions

Whereas some quantities cannot be reduced in their number due to different measurands, the physical units can be reduced by convention down to a single physical unit like the second. The SI base units are dependent by their definitions containing the factors mutually adjusted by the CODATA. The metre is dependent on the second because the length cannot be measured without using time. However, the light-second, for example, is not a practical unit for the length. This is the well-known problem of any natural units. Thus, some practical "expansion" of the SI system using a non-unitary factor is useful. However, it is not the case of the radian with the unitary factor. In [1] the author only believes (coining its own definition, not by a proof) that the plane angle has a physically independent dimension (i.e. e.g. as the third independent dimension for $\theta$ in a 2D plane *XY*) contrary to the well-established practice in mathematics, to the SI brochure as well as to the VIM. The unit m/m = 1 of the plane angles may not be necessarily have a special name, the radian that simply represents 1. The proposed redefinitions of the radian and the steradian are not based on any physical experiment (i.e. it cannot be checked independently). Thus, *the radian should be neither a base unit nor a complementary unit and it is not necessary for it to be redefined*. The "Radian Convention" is acceptable, and thus also no reformulations of equations are needed. However, it is not necessary to be explicit, as it is suggested in [1]. We do not generally need an explicit statement that we are omitting the number one in all multiplications, as in the case also for the radian. It seems that any *pure quantity equation* with the plane angle is not presented in the author's papers as a correct one, whereas *the units must be generally hidden* in quantity equations.

The wording in the ISO definition of the frequency and dependent quantities should be general to deal also with non-harmonic frequencies and the use of the reciprocal second "$s^{-1}$" should not be restricted (as well as the reciprocal metre "$m^{-1}$" is not restricted). The becquerel (Bq = $s^{-1}$) that is restricted to stochastic processes of radionuclides is not a sufficient alternative. The quantity equation defining the *angular frequency will remain the same* equation $\omega = 2\pi\nu$ (constant $\theta_N$ will not be introduced at all), its name can be changed to "the radian frequency" or "the frequency of radians", i.e. a special case of "the frequency of events" with a rotation by a radian as an event. However, this is not necessarily due to calculations because it is only a language-dependent convention.

The *dimension for the plane angle unit cannot be accepted* because different time-measurement principles will lead to various units for time. If the plane angle measurement is, for example, involved in the measurement of time (e.g. for the timescale UT based on the Earth's rotation), then the radian with its own dimension cannot be lost in the result connected through a measurement model that is also used for the uncertainty calculations. The result of the time measurement will have a unit like "s/rad", with only one exception. If the radian will be divided by another radian (i.e. rad/rad), then information will be lost. However, this is analogous to an information loss that the author of [1] intended to avoid from the beginning in the case of the "m/m" ratio.

In [1], the author's proposals are very unconventional and they do not seem to be sufficiently thought out. They will break historical continuity of the metric system and bring many changes in the whole physics due to many redefinitions of quantities based on the rewritten quantity equations due to unhiding one unit that is conventionally equal to one. The "profound step for the SI" needs a strong support. However, the necessity is insufficiently presented. A better solution for an easier usability of the SI is to remove restrictions rather than to add extra things.